\documentclass[11pt]{article}

\usepackage{fullpage}
\usepackage{amsmath}
\usepackage{amssymb}
\usepackage{setspace}
\usepackage{bbm}
\usepackage{dsfont}
\usepackage{graphics}

\usepackage{color}

\usepackage[colorlinks=true]{hyperref} 
\hypersetup{
    bookmarks=true,         % show bookmarks bar?
    unicode=false,          % non-Latin characters 
    pdftoolbar=true,        % show Acrobat
    pdfmenubar=true,        % show Acrobat 
    pdffitwindow=false,     % window fit to page when opened
    pdfstartview={FitH},    % fits the width of the page to the window
    pdftitle={My title},    % title
    pdfauthor={Author},     % author
    pdfsubject={Subject},   % subject of the document
    pdfcreator={Creator},   % creator of the document
    pdfproducer={Producer}, % producer of the document
    pdfkeywords={keyword1} {key2} {key3}, % list of keywords
    pdfnewwindow=true,      % links in new window
    colorlinks=true,       % false: boxed links; true: colored links
    linkcolor=blue,          % color of internal links (change box color with linkbordercolor)
    citecolor=red,        % color of links to bibliography
    filecolor=magenta,      % color of file links
    urlcolor=cyan           % color of external links
}

\usepackage[normalem]{ulem}
\usepackage{amsmath}
\usepackage{enumerate}
\usepackage{amsfonts}
\usepackage{yfonts}

\usepackage{subfigure}
\usepackage{psfrag}

\usepackage{epsfig}
\usepackage[latin1]{inputenc}
\usepackage{float}
\usepackage{graphicx}
\usepackage{cancel}
\usepackage{mathrsfs}
\usepackage{amssymb}
\usepackage{amsfonts}
\usepackage{amsmath}
\usepackage{slashed}
\usepackage[font=small,labelfont=bf]{caption}

\newcommand \tr {\mbox{{\bf Tr}}}

\usepackage{graphicx}
\usepackage{bm}

\newcommand{\be}{\begin{equation}}
\newcommand{\ee}{\end{equation}}
\newcommand{\bes}{\begin{equation*}}
\newcommand{\ees}{\end{equation*}}
\newcommand{\bea}{\begin{eqnarray}}
\newcommand{\eea}{\end{eqnarray}}
\newcommand{\beas}{\begin{eqnarray*}}
\newcommand{\eeas}{\end{eqnarray*}}

\newcommand{\p}{\partial}

\newcommand{\bmat}{\begin{bmatrix}}
\newcommand{\emat}{\end{bmatrix}}

\newcommand{\ZZ}{\mathbb{Z}}
\def\tr{{\rm tr}}

\def\le{\left}
\def\ri{\right}

\begin{document}

\numberwithin{equation}{section}
{
\begin{titlepage}
\begin{center}

\hfill \\
\hfill \\
\vskip 0.75in

{\Large \bf Fake gaps in AdS$_3$/CFT$_2$}\\

\vskip 0.4in

{\large Alexandre Belin${}^a$, Alejandra Castro${}^b$ and Ling-Yan Hung${}^{c,d}$}\\

\vskip 0.3in

{\it ${}^{a}$ Department of Physics, Stanford University, \\
Stanford, CA 94305-4060, USA} \vskip .5mm   
{\it ${}^{b}$ Institute for Theoretical Physics, University of Amsterdam,\\
Science Park 904, Postbus 94485, 1090 GL Amsterdam, The Netherlands} \vskip .5mm    
{\it ${}^{c}$ Department of Physics and Center for Field Theory and Particle Physics, Fudan University, \\
220 Handan Road, 200433 Shanghai, China} \vskip .5mm
{\it ${}^{d}$ Collaborative Innovation Center of Advanced  Microstructures,
Fudan University,\\ 220 Handan Road,  200433 Shanghai, China.}

\end{center}

\vskip 0.35in

\begin{center} {\bf ABSTRACT } \end{center}
We discuss properties of interpolating geometries in three dimensional gravity in the presence of a chiral anomaly. This anomaly, which introduces an unbalance between left and right central charges, is protected under RG flows. For this simple reason it is impossible to gap a system with such an anomaly. Our goal is to discuss how holography captures this basic and robust feature. We demonstrate the absence of a mass gap by analysing the linearized spectrum and holographic entanglement entropy of these backgrounds in the context of AdS$_3$/CFT$_2$.

\vfill

\noindent \today

\end{titlepage}
}
%%%%%%%%%%%%%%%%%%%%%%%%%%%%%%%%%%%%%%%%%%%%%%%%%%%%%%%%%%%%%%%%%%

%%%%%%%%%%%%%%%%%%%%%
\newpage

\tableofcontents
\newpage

\section{Introduction }

One ambitious question that has long confounded field theorists is the classification of possible IR behaviour for a given UV fixed point subjected to arbitrary relevant perturbations. A, C and F theorems, whose various variations proved in 2d, 3d and 4d constitutes a powerful set of constraints on possible renormalization group (RG) flows in Lorentz invariant and unitary theories. The most notable success being Zamolodchikov's theorem \cite{Zamolodchikov:1986gt}, and the recent proof of Komargodski-Schwimmer \cite{Komargodski:2011vj, Komargodski:2011xv}.  

The AdS/CFT correspondence has brought many new insights in understanding this classification. One radical  feature in the correspondence is that different characteristics and phenomena in a quantum field theory---such as RG flows or entanglement entropy between different regions---manifest themselves geometrically in the dual gravity description. 
Constraints on RG flows have been studied exentsively holographically\footnote{There is an overwhelming amount of literature on this subject. We refer the reader to \cite{Freedman:1999gp,Girardello:1998pd,Bianchi:2001de} for early work on the subject; see \cite{Myers:2010xs,Myers:2010tj} and references within for a more current status on the subject.} and insights learned in this process, such as the connection to entanglement entropy \cite{Casini:2011kv}, has been crucial in understanding the above theorems. In particular, its influence in the three dimensional F-theorem  \cite{Casini:2012ei,Jafferis:2011zi,Klebanov:2011gs} is without doubt.

There are other features that constrain RG flows in quantum field theory. Most notably, quantum anomalies play a significant role. For example,  it was argued by 't Hooft that  ABJ type anomalies of continuous global symmetry groups  are necessarily conserved along  a RG flow, i.e. the UV and IR theory suffer from the same amount of anomaly \cite{tHooft:1979bh}. Very generally, quantum anomalies have shown up as the most important defining signature of the boundary states of different phases of matter found in condensed matter, most notably the classes of state named ``symmetry protected topological" (SPT) phase \cite{cglw,Wen:2013oza,KongWen}. In that context, it is also argued from very general grounds that boundaries of these exotic phases are robust against developing mass gap via adding any symmetry preserving perturbations in the infrared \cite{cglw,Luvishwanath,senthilwang}.

 Even though it is understood how to design a holographic system that can flow to a gapped system in the IR, it has not been established how quantum anomalies are robust and protected geometrically.  It is thus a curiosity how such ``un-gappability" presents itself geometrically, where one might expect some restrictions over admissible background solutions. Our aim is to capture the ``un-gappability" holographically and quantify how geometry knows about the powerful theorems for quantum anomalies.

The current paper is a step toward extracting a geometric signature of quantum anomalies.  While it would be interesting to study similar effects in higher dimensions, the discussion here will be limited to two dimensional QFTs and their three dimensional holographic duals.           

\subsection{Chiral anomalies}

The simplest and most elegant example in field theory where a protected anomaly arises is  the chiral anomaly in 1+1 dimensions. This is a theory where the OPE of the stress tensor is given by 
\bea
T(w)T(0) &\sim& {c_L/2\over w^4} + {T(0)\over w^2}+{\p T(0)\over w} + \ldots~,\cr  && \cr
\bar T(\bar w)\bar T(0) &\sim& {c_R/2\over \bar w^4} + {\bar T(0)\over \bar w^2}+{\bar{\p}\bar T(0)\over \bar w} + \ldots~,
\eea
with $\bar{T}(\bar w)\equiv 2\pi T_{\bar w\bar w}$  and $T(w)\equiv 2\pi T_{ww}$; moreover, its chiral nature comes from having $\hat c =c_L-c_R \neq0$. $c_L$ and $c_R$ are commonly denoted the left and right central charges. Whereas $c=(c_L+c_R)/2$ can flow according to Zamolodchikov's theorem, it is also well known that $\hat c \neq0$ is protected.\footnote{A simple argument goes as follows: modular invariance, in particular $T$ transformations, imposes that $ 24 \hat c \in {\ZZ}$, and hence $\hat c$ cannot continuously vary. An argument that is more general goes along the lines of the analysis presented in section \ref{sec:u1} for conserved currents. See also \cite{Bastianelli:1996gh,Hotta:2009zn} for a complete derivation. } And for this simple reason, adding any symmetry preserving perturbation will not generate a  mass gap in the infrared.

The chiral anomaly provides a robust constraint on RG flows. But, {\it how does the 2+1 holographic dual theory know about this constraint?} Examples of gravity duals of chiral CFTs are already known, such as, most notably, topologically massive gravity theory in three dimensions (TMG) \cite{Deser:1981wh,Deser:1982vy,Deser:1991qk}. This is thus an ideal playground to observe how this protected anomaly plays out in the bulk space-time.

The relevant comparison in holography is to contrast results of Einstein gravity coupled to matter versus TMG coupled to matter. In Einstein's theory it is possible to craft geometries whose  UV structure is conformal and the IR behaviour can be designed to be gapped.  The surprise is that these backgrounds persist to be {\it exact} classical solutions in TMG! It must then be the case that, even though the background is unchanged, observables change dramatically after including a chiral anomaly in the bulk three dimensional theory. 

Given the expectation that the boundary is robust against arbitrary relevant perturbations, we will study perturbations around a generic RG geometry preserving boundary Lorentz invariance sourced by a scalar field with very general interaction potentials $V(\phi)$, and demonstrate how the boundary CFT stays gapless all the way. This is in sharp contrast to the case of Einstein theory where one readily finds holographic RG flows leading to a gapped infrared theory. We will also confirm such expectation by studying holographic entanglement entropy.  The standard Ryu-Takayanagi prescription \cite{Ryu:2006bv,Ryu:2006ef} is modified by this anomaly, which was developed in \cite{Castro:2014tta}, and its modification is crucial to account for the expectation from the dual boundary theory.  This is a non-trivial test of the proposal in  \cite{Castro:2014tta}, since the success shown here does not rely on symmetries of the background: it is a dynamical feature of the probe used to evaluate entanglement entropy. The effects of gravitational anomalies on entanglement entropy, on both sides of the holographic duality, have been recently generalized to higher dimensions in \cite{Guo:2015uqa,Azeyanagi:2015uoa,Nishioka,IqbalWall}. 

Another approach would be to build holographic $c$-functions where the energy scale is parametrized by the radial direction. This was the approach used in  \cite{Hotta:2009zn}, and they showed that their  was no radial dependence on $\hat c$. In our opinion, this approach carries some ambiguities since there is no robust definition of such $c$-functions (or at least the precise interpretation of it).  Our derivations and conclusions do not rely on these definitions, nevertheless we are in perfect agreement with the results reported in \cite{Hotta:2009zn}.

\subsection{Outline}

This paper is organized as follows. In section \ref{sec:general} we will review the criteria that allows for a given bulk  solution to be interpreted as a RG flow that interpolates between a conformal UV fixed point and a gapped theory in the IR. The two key observables that we will use to signal a gapped IR are linearized spectrum and holographic entanglement entropy. The discussion in section \ref{sec:general} is valid for any dimension, but in later sections we will limit to three dimensional bulk theories.  In section \ref{sec:tmg}  we introduce a chiral anomaly by including a gravitational Chern-Simons term in three dimensions. For generic non-AdS backgrounds, we will show that the IR physics are dominated by the gravitational anomaly. This is clearly reflected by studying both linearized perturbations and holographic entanglement entropy. In the first two appendices we cover content related to conventions and details on the results related to section \ref{sec:tmg}.  
And in appendix \ref{sec:u1} we discuss how similar results holds for holographic theories that have unbalanced gauge anomalies. In this case the effects of a quantum anomaly on the RG flow is not reflected classically in the bulk geometry; instead we find that quantum effects in the gravitational dual have to be included to appropriately capture the  ``un-gappability" of the system.

\section{Gapped holographic systems: entanglement entropy and spectral analysis}\label{sec:general}

The class of holographic systems we will study are vacuum (i.e. zero energy, zero entropy) solutions which in the UV are conformal and the interior IR geometry has a non-trivial radial profile. Depending on this radial profile, one could infer that the IR geometry is dual of a QFT with a gapped spectrum. There are two common holographic routes to determine if the IR geometry is gapped or not.\footnote{There are of course more ways to test if the IR point is gapped. For instance, the thermodynamics response of the system; see \cite{Jensen:2012kj,Jensen:2013kka}.}
The first route is to explicitly quantify the spectrum by studying linearized perturbations around the given background of interest, see for example \cite{Bianchi:2001de,Charmousis:2010zz,Liu:2013una}. The second route is to study the behaviour of the entanglement entropy, as in \cite{Klebanov:2007ws,Myers:2012ed,Liu:2013una}. These are the two observables (spectrum and entanglement) that we will discuss throughout this work. In this section we will summarize the main features of these observables when the dynamics of the holographic background  is governed by an Einstein-Hilbert action coupled to matter. 

Our starting point is to consider geometries which we parametrise as
\be\label{eq:zg}
ds^2 =\frac{\eta_{ij} dx^i dx^j}{z^2} + \frac{dz^2}{z^2 f(z)}~,\quad i=0,\ldots, d~.
\ee
As $z\to0$, i.e. in the UV limit, the spacetime  asymptotes to AdS$_{d+1}$ which requires $f(z)\to 1$. In the infrared limit, $z\to \infty$, we will take $f(z) \sim z^n$ for some power $n\geq0$.
This geometry can be a solution of an Einstein-Scalar system 
\be\label{action}
I[g,\phi] = \frac{1}{16\pi G_3} \int d^{d+1}x \sqrt{g} \left(R - 8( \nabla\phi)^2 - V(\phi)\right)~.
\ee
%where the scalar field adopts a potential $V(\phi)$ 
%\footnote{Note that here the kinetic term of the scalar is not normalized as in \cite{Charmousis:2010zz}. We use a convention where the kinetic term in the action is $\frac{1}{16\pi G_3}\int d^3x \sqrt{g} \le( 8 (\nabla \phi)^2+\cdots\ri)$.}. 
Provided $f(z) \sim z^n$, it is sufficient to have 
\be
V(\phi) \sim \exp(- 4\delta \phi)~,
\ee
in which case the power $n$ can be expressed in terms of $\delta$ as
\be
n = (d-1) \delta^2~.
\ee
The null energy condition requires $n<2d$.  In the following we will determine for which ranges of $n$ the infrared geometry can be interpreted as dual to a gapped system.  

Before proceeding it is important to note that these solutions are singular in the interior: as $z\to \infty$ there is a curvature singularity in \eqref{eq:zg} and the scalar field diverges. This is an apealing feature since it makes evident that the background is not a black hole (i.e. a thermal state), and hence it can be interpreted as the RG interpolation between two theories. For our purposes the singular nature of the geometry does not spoil the analysis: what is important is that we can properly quantify the observables which will be evident below. Moreover, for $0\leq n <2d$, it has been argued extensively that the singularity can be resolved either by embedding into string theory or other methods \cite{Charmousis:2010zz,Gubser:2000nd}.

%Considering the transverse traceless part of the background perturbation, it is found that the metric perturbation $h_{\mu\nu} = z^{-(d-1)/2} \psi \epsilon_{\mu\nu}(k,\omega) \exp(- i (\omega t - k.x)) $ satisfies the equation
The simplest way to quantify the spectrum, is to consider a massless probe scalar $\Psi(x^\mu)$ on the background  \eqref{eq:zg}. The resulting Schrodinger equation is
\be
\partial_u^2{\psi} + V_{\rm eff}(z(u)) \psi = k^2\psi~,\quad V_{\rm eff}(z)={d-1\over 2}\sqrt{z^{d-1} f(z)}{\partial\over\partial z}\le( \sqrt{f(z)\over z^{d+1}}\,\ri)~,
\ee
where  
\be
\Psi(z,x^\mu) = z^{(d-1)/2} e^{i k_i x^i} \psi(z)~, \quad  \partial_u\psi \equiv \sqrt {f(z)} \partial_z \psi~,
\ee
and $k^2\equiv\eta_{ij}k^i k^j$; the radial coordinate $u$ is defined appendix \ref{app:cn}. In the IR limit, we have $f(z) \sim z^n$ and the effective potential becomes
\be\label{eq:veff}
V_{\rm eff} \sim z^{n-2} .
\ee
For $n>2$, as $z\to \infty$, the potential $V_{\rm eff}$ diverges.
Since $V_{\rm eff}$ also diverges in the UV limit, this suggests that we have an infinite well, and
the spectrum is discrete in this well. Therefore, there should be a mass gap.
For $n<2$ however, $V_{\rm eff}$ vanishes as $z\to \infty$, suggesting that it
is a half well, and the spectrum is continuous and hence ungapped. The case $n=2$ requires more care, and we refer to \cite{Charmousis:2010zz,Liu:2013una} for a complete discussion. 

This simple example of a probe scalar gives us a range of $n$ for when one should expect a gapped spectrum  ($2\leq n<2d$) or not ($0\leq n< 2$), and for many cases this probe is enough. However, it will be evident in later sections that probe fields are blind to the true nature of the system since they might not capture key aspects of the spectrum.
 A proper treatment requires studying the actual metric and background fields perturbations as done in e.g. \cite{Bianchi:2001de,Charmousis:2010zz}. In appendix \ref{app:EES} we do this explicitly for a three dimensional version of \eqref{action}.

\begin{figure}[ch]
\begin{center}
\includegraphics[width=0.7\textwidth]{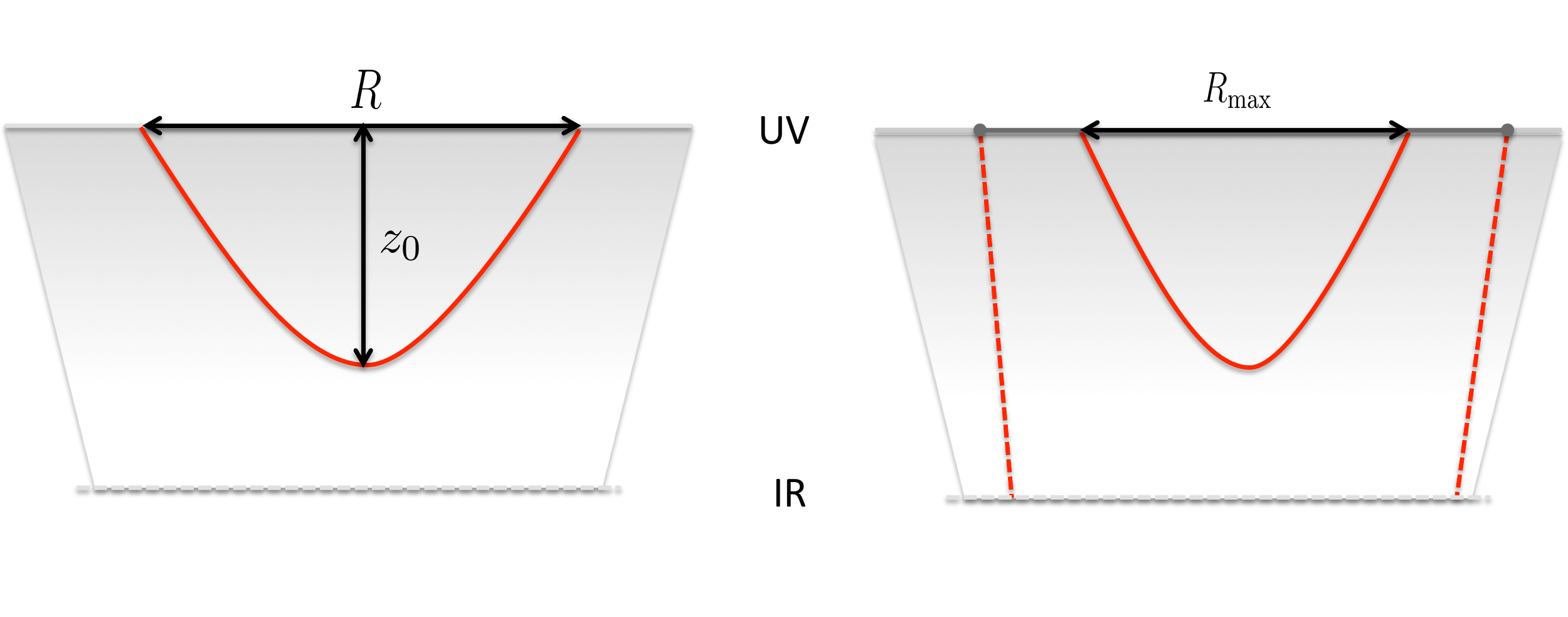}
\end{center}
\caption{Schematic representation on the behaviour of minimal surfaces for a geometry that is not gapped (left picture) contrasted to a gapped background (right picture). The solid red lines correspond to connected surfaces, which are anchored at the boundary. The dotted red lines are disconnected surface, which dominates $S_{\rm EE}$ for $R>R_{\rm max}$.}\label{fig:1}
\label{fig:boost}
\end{figure}

One could alternatively use holographic entanglement entropy as a probe of the spectrum: the topology of the entangling surface changes dramatically depending on the value of $n$, which is a signal of a gap in the system \cite{Klebanov:2007ws,Myers:2012ed,Liu:2013una}. The argument  can be summarised as follows. Consider  the relation between the length of a strip at the boundary, which we denote $R$,  as a function of the turning point of the minimal surface, denoted $z_0$. For the background \eqref{eq:zg} this would give
\be\label{eq:hlmm}
R= 2\int_0^{z_0} {dz\over f(z)^{1/2}} {z^{d-1}\over \sqrt{z^{2(d-1)}_0- z^{2(d-1)}}}~.
\ee  
For large $z_0$ we can replace $f(z)\sim f_0 z^n$, and we will have 
\be
R\sim 2z_0^{1-n/2}(\chi f_0^{-1/2} + O(z_0^{-1}))~,\quad \chi = \int_0^1 {u^{1-n/2}du\over \sqrt{1-u^2}}~,
\ee
while for small $z_0$, the leading effect is given by $f(z)\sim 1$, and we get
\be\label{eq:hlmm1}
R\sim 2 z_0 + O(z_0)~.
\ee
What these manipulations clearly illustrate is that if $n>2$ then $R\to 0$ both as $z_0$ increases and decreases. This indicates that connected geodesics have  a maximal value $R^{\rm max}$ when $n>2$, and hence for boundary intervals with $R>R^{\rm max}$ 
the disconnected solution would take over. This is illustrated in figure \ref{fig:1}.  The absence of $R$ in  $S_{\rm EE}$ for disconnected geodesics signals that the system is  gapped in the IR. The analysis and bounds on $n$ using entanglement is in complete agreement with that of the spectral analysis reviewed above.\footnote{For $d=2$ and $n=1$ there is a small caveat to this line of reasoning. There are situations where the connected solution is always dominant, but as  $R\to\infty$ the length of the curve does not depend on $R$. This is simply due to the fact that the effective central charge asymptotes to zero in the IR, and it could be interpreted as a signal of a ``marginal gap'' in the spectrum. In this context marginal refers to the deformation inducing a very rapid power law decay of observables with distance, rather than exponential decay (which we would denote as a ``hard'' gap). }

\section{Chiral anomalies in AdS$_3$/CFT$_2$}\label{sec:tmg}

%Provide some basic properties of gapped systems studied in prior sections, we would like to investigate which of these features are preserved or not in the presence of the topological mass term $S_{TMG}$ in addition to Einstein-Hilbert term. 
A simple way to prevent a gap in the IR is to introduce a protected anomaly. In a two dimensional CFT this can be done by adding a chiral anomaly, i.e.  by having an unbalance in left versus right central charge, $c_L\neq c_R$. Whereas the total central charge $c=(c_L+ c_R)/2$ is a monotonic function in the space of RG flows, the difference $\hat c=c_L-c_R$ is  protected: given a value of $\hat c$ in the UV, it remains intact in the IR. In this section we will show how three dimensional gravity protects $\hat c$.

%It is then rather clear that if $\hat c \neq 0$ in the UV, the IR fixed point is still conformal.\footnote{Depending of the deformation in the UV, one could flow to an IR theory that has  $c\to 0$ and  $\hat c$ fixed. In this case either $c_L$ or $c_R$ is negative in the IR, which indicates that the theory is not unitary. We will see that this is what usually happens in holography. {\color{red} AB - do we really see/show this?}} {\color{red} AB - do you mean that there must be an IR fixed point?}

Holographically it is known how to induce a non vanishing value of $\hat c$ in three dimensional gravity. Such systems are described by the action principle \eqref{action} with the addition of a gravitational Chern-Simons term
\be
I_{\rm GCS} = \frac{1}{32\pi G_3 \mu}\int d^3x \sqrt{g} \epsilon^{\lambda\mu\nu} \left(\Gamma^\rho_{\lambda \sigma} 
\partial_\mu \Gamma^\sigma_{\rho\nu} + \frac{2}{3} \Gamma^\rho_{\lambda\sigma}
\Gamma^\sigma_{\mu\tau}\Gamma^\tau_{\nu\rho}\right) ~.
\ee
Gravitational theories that include both the Einstein-Hilbert term and $I_{\rm GCS}$ are known as topologically massive gravity (TMG) \cite{Deser:1981wh,Deser:1982vy,Deser:1991qk}. The role of this term is to induce a gravitational anomaly. In particular for a space-time that is asymptotically AdS$_3$  the gravitational Chern-Simons term induces an unbalance in  the central charges given by \cite{Solodukhin:2005ns,Kraus:2005zm,Hotta:2008yq,Skenderis:2009nt} 
\be
c_L = {3\ell \over 2G_3}(1+ {1\over \mu \ell})~,\quad c_R = {3\ell \over 2G_3}(1- {1\over \mu \ell})~.
\ee
 Our goal is to show that for finite  $\mu$ it is impossible to have a background solution that could be interpreted in the IR as a gapped system.  

One might think that the obstruction in gapping the IR geometry would be to argue that a ``gapped'' background ceases to exist  after including  $I_{\rm GCS}$. This would have been a clean and elegant resolution,  however, it is not the case.
The contribution of $I_{\rm GCS}$ to the equations of motion is given by
\bea\label{eq:EHS1}
R_{\mu\nu} -{1\over 2} g_{\mu\nu} R+{1\over \mu} C_{\mu\nu}= 8 \partial_\mu \phi \partial_\nu \phi - 4 g_{\mu\nu} \partial^\alpha \phi \partial_\alpha \phi - {1\over 2} g_{\mu\nu} V(\phi)~.
\eea
where $C_{\mu\nu}$ is the Cotton tensor:
\be
C_{\mu\nu} = \epsilon_\mu^{\,\,\,\,\alpha\beta} 
\nabla_{\alpha}(R_{\beta\nu} - \frac{1}{4} R g_{\beta\nu}) = \frac{1}{2}(\epsilon_\mu^{\,\,\,\rho\sigma}
\nabla_{\rho}R_{\sigma\nu} + \epsilon_\nu^{\,\,\,\rho\sigma}
\nabla_{\rho}R_{\sigma\mu})~.
\ee
Despite the complexity of this equation, it was noted in \cite{Hotta:2009zn} that for all metrics of the  form 
\bea\label{eq:geng}
ds^2 &=& \frac{dz^2}{z^2 f(z)}+  \frac{\eta_{ij} dx^i dx^j}{z^2}\cr
&=&{1\over z(u)^2}\le( du^2 +  \eta_{ij} dx^i dx^j\ri)~,
\eea
the Cotton tensor vanishes for arbitrary $z(u)$. This is simply due to the fact that the metric \eqref{eq:geng} is conformally flat and thus  $C_{\mu\nu}=0$.  This means that if \eqref{eq:geng} is a solution of a  two derivative action, it  would automatically be a solution after adding $I_{\rm GCS}$. All backgrounds discussed in section \ref{sec:general} fall into this category, and hence it remains a solution even when $c_L\neq c_R$.  Moreover, using the prescription of  e.g. \cite{Hotta:2009zn,Myers:2010xs,Myers:2010tj}, the holographic $c$-function is given by
\be\label{eq:cz}
c(z)={ 3\over 2G_3 f(z)^{1/2}}~,
\ee
which is unaffected by the gravitational Chern-Simons term since it is built out of traces of the equations of motion. This quantity suggests strongly the total central charge $c=c_L+c_R$ is decreasing monotonically along the radial `RG' direction. But this is of course not enough to infer what happens to $c_L(z)$ and $c_R(z)$ independently; moreover the dual interpretation of \eqref{eq:cz} far from the fixed points is somewhat ambiguous. Still this raises some concerns: if we naively think that the results of section \ref{sec:general} still hold we would interpret  \eqref{eq:geng} as a gapped system in the IR for a theory with $c_L\neq c_R$. But this cannot be true! We just argued that it is impossible to gap a UV system with unbalanced left versus right moving excitations. Holography seems to say the opposite. In the following we will resolve this puzzle by quantifying the linear bulk modes and holographic entanglement entropy in the presence of the gravitational Chern-Simons term.

\subsection{Linearized spectrum}\label{sec:tmgpert}

Our first step towards resolving our puzzle is to look at the linearized perturbations of the matter fields that support the background \eqref{eq:geng}. The difficult portion is encapsulated in the Cotton tensor in \eqref{eq:EHS1}. Prior literature on linearising $C_{\mu\nu}$ includes e.g. \cite{Deser:2002iw,Li:2008dq, Skenderis:2009nt}. Within this literature, it is very well known that TMG has negative energy excitations, unless we are at the chiral point  \cite{Li:2008yz,Grumiller:2008qz,Maloney:2009ck}. This feature will still be present here, however we will not highlight this explicitly below. Our aim is to quantify if the spectrum is discrete or continuos in the IR geometry.%, and we leave a detailed discussion of unitarity issues for future work.

For the purpose of describing the linearized equations we will use a radial gauge
\be
g_{\mu\nu}= g^{(0)}_{\mu\nu} +  h_{\mu\nu} ~,\quad \phi = \phi_{(0)} + \delta\phi~, \quad  h_{u\mu}=0~,
\ee
where  $(g^{(0)}_{\mu\nu}, \phi_{(0)})$ define the background solution, and $(h_{\mu\nu},\delta\phi)$ are the small perturbations around the background. Details of the computation are presented in appendix \ref{app:1}. It is convenient to write the background solution \eqref{eq:geng} as
\be\label{eq:uu}
ds^2 =  g^{(0)}_{\mu\nu} dx^\mu dx^\nu=e^{2A(u)}\le(du^2 +  \eta_{ij} dx^i dx^j\ri)~, \quad e^{-A(u)}\equiv z(u).
\ee
and we will use null coordinates for the boundary directions: 
\be
w=-t + x ~,\quad \bar w = t+x ~, \quad \partial \equiv\partial_w~,\quad \bar\partial\equiv \partial_{\bar w}~.
\ee 
The background scalar $\phi_0$ satisfies \eqref{eq:ap1}-\eqref{eq:ap2}.

The differential equations for $h_{\mu\nu}$ and $\delta \phi$ are 
\bea\label{eq:yi}
{1\over \mu}\bigg(\partial_u^2 \bar\partial  h_{w w} -{2}\partial^2 \bar\partial h_{w\bar w} +\partial \bar\partial^2 h_{w w} +\partial^3  h_{\bar w\bar w} \bigg)- e^A \partial_u\partial  h_{w\bar w}
+ e^A\partial_u \bar\partial h_{w w} -8e^A\dot\phi_{(0)}\,\partial \delta \phi=0~,\cr
{1\over \mu}\bigg(\partial_u^2\partial  h_{\bar w \bar w} -2\partial\bar\partial^2 h_{w\bar w} + \bar\partial^3 h_{w w} +\partial^2\bar\partial  h_{\bar w\bar w}\bigg) - e^A\partial_u\partial  h_{\bar w\bar w}
+ e^A\partial_u\bar\partial  h_{w \bar w}+8e^A\dot\phi_{(0)}\,\bar\partial \delta \phi=0~,
\eea
and
\be\label{eq:trace}
\partial_u\le( 8e^{A}\dot\phi_{(0)}\delta  \phi+ e^A\partial_u  h_{w\bar w}  +{1\over \mu}\big(\bar\partial^2 h_{ w w}-\partial^2 h_{\bar w\bar w}\big)\ri)=0~.
\ee
These equations correspond to the $(i,j)$ components of \eqref{eq:EHS1}, and they take into account constraints conditions which arise from the $(u,i)$ components in \eqref{eq:EHS1}. In the above we defined $\dot\phi_0\equiv \partial_u\phi_0$ and $\dot A \equiv \partial_u A(u)$.
There is also a constraint which is obtained by tracing \eqref{eq:EHS1} and the linearized equation of motion for $\phi$. These equations can be found in appendix \ref{app:1}.

It is important to note that all terms multiplying $\mu^{-1}$ in the above equations are independent of the conformal factor $e^{2A(u)}$: this is simply because the Cotton tensor is invariant under Weyl transformations and our choice of coordinates makes this explicit. This will be key in the following subsections. 

Finally, it is possible to manipulate the above equations such that $(h_{w\bar w},\delta \phi)$ essentially decouple from $(h_{ww},h_{\bar w\bar w})$, and hence $(h_{w\bar w},\delta \phi)$ should be taken as  source terms in the equations for $h_{ww}$ and $h_{\bar w\bar w}$. The decoupled equations governing $h_{w\bar w}$ and $\delta \phi$ are given in the appendix \ref{app:deq}. While this decoupling is conceptually important, the details of the equations are not crucial and we refer the reader to the appendix for details. 

\subsubsection{IR limit: conformal gravity}\label{sec:pir}

Our aim is to understand the behavior of the modes $h_{\mu\nu}$ and $\delta \phi$ in the IR geometry. Along the lines of the discussion surrounding \eqref{eq:veff}, the behavior of the modes in the interior dictates if we have bound states or not. 

In the notation \eqref{eq:zg}, the IR geometry is given by $z\to \infty$ and $f(z)\sim z^n$ with $0<n<4$ (the upper bound for $n$ is dictated by the null energy condition). In terms of the radial coordinate $u$ in \eqref{eq:uu}, the IR geometry is given by
\bea\label{eq:uunear}
&u\sim z^{-n/2+1} \to \infty ~\quad& {\rm if}~ 0<n<2~,\cr
&u\sim \log(z) \to \infty ~\quad& {\rm if}~ n=2~,\cr
&u\sim u_0+ z^{-n/2+1} \to u_0 ~\quad& {\rm if}~ 2<n<4~,
\eea
where $u_0$ is a positive constant.\footnote{And for completeness, the UV boundary is located at $u\sim z \to 0$ with $f(z)\sim 1$.} In addition we have $e^{-A(u)}=z$ which diverges in the interior. And according to \eqref{eq:ap1}, the background scalar behaves in the IR as $e^A\dot \phi_{(0)} \sim z^{n/2-2} $ which decays as $z\to\infty$ for the range of $n$ we are considering. 

 Zooming into the IR geometry, we see that dominant terms in equations \eqref{eq:yi}-\eqref{eq:trace} are 
\bea\label{eq:cgl}
{1\over \mu}\bigg(\partial_u^2 \bar\partial  h_{w w} -{2}\partial^2 \bar\partial h_{w\bar w} +\partial \bar\partial^2 h_{w w} +\partial^3  h_{\bar w\bar w} \bigg)&=&0\cr
{1\over \mu}\bigg(\partial_u^2\partial  h_{\bar w \bar w} -2\partial\bar\partial^2 h_{w\bar w} + \bar\partial^3 h_{w w} +\partial^2\bar\partial  h_{\bar w\bar w}\bigg)&=&0\cr 
{1\over \mu}\partial_u\bigg(\bar\partial^2 h_{ w w}-\partial^2 h_{\bar w\bar w}\bigg)&=&0
\eea
In this limit we kept  $(\partial, \bar \partial)$ fixed and we discarded terms like $e^A\partial_u$ since they are subleading relative to $\partial^2_u$. These equations are those relevant for conformal gravity, i.e. a gravitational theory described solely by $I_{\rm GCS}$.\footnote{For a recent discussion on conformal gravity and a complete set of references see \cite{Afshar:2011qw,Afshar:2011yh}.} The Cotton tensor is insensitive to $e^{A(u)}$ and dominant in the IR: this is already indicating that the behavior of the modes is independent of $n$ which should be contrasted with the discussion in section \ref{sec:general} and appendix \ref{app:EES}. This is  a rather powerful observation that has no counterpart in the absence of the gravitational anomaly. 

Let us solve explicitly \eqref{eq:cgl}. It is important to note that $h_{w\bar w}$ contributes as a non-homogenous term for $(h_{ww},h_{\bar w\bar w})$ since we can solve for  $h_{w\bar w}$ using the equations in appendix \ref{app:deq}. The homogenous solutions to \eqref{eq:cgl}  are simply 
%
%equations for the other components are
%
%\be
%\partial_u^2  h_{w w}  +2 \partial \bar\partial h_{w w} =0~, \quad \partial_u^2  h_{\bar w \bar w}  +2 \partial \bar\partial h_{\bar w \bar w} =0~.
%\ee
%The solutions are  plane waves and in momenta space we have
\be\label{eq:hpw}
h_{ww}\sim e^{\pm i\sqrt{2\partial \bar \partial}\, u}~,\quad h_{\bar w\bar w}\sim e^{\pm i\sqrt{2\partial \bar \partial}\, u}~,
\ee
which holds regardless of the details of the properties of the radial profile. In particular, both solutions are regular in the interior so we conclude that these modes have a continuous spectrum; moreover the details of the IR geometry are irrelevant. In appendix \ref{app:deq} we discuss the IR behavior of $\delta \phi$ and $h_{w\bar w}$; in contrast to \eqref{eq:hpw} these modes are sensitive to the details of the IR geometry and their spectral functions can be gapped. But since there is a sector of the theory that is always continous, which \eqref{eq:hpw} makes evident, we conclude that the whole system is {\it not} gapped in the IR.  

\subsection{Entanglement entropy}

We have just shown that, despite the geometric properties of the background \eqref{eq:geng}, in TMG  {\it all} vacuum solutions have a continuous spectrum in the IR region. In the following we would like to see how this is captured by entanglement entropy. This will prove how physical observables are starkly different even though the background geometry is unchanged. 

In the presence of a gravitational anomaly, the Ryu-Takayanagi prescription \cite{Ryu:2006bv,Ryu:2006ef} is modified. The modification due to the gravitational Chern-Simons term can be interpreted as the functional of a massive spinning particle \cite{Castro:2014tta}. Holographic entanglement entropy is given by the minimum of the functional\footnote{For any vector $V^\mu$, we define, 
$$
\nabla V^\mu\equiv {d V^\mu\over d \tau}+\Gamma^{\mu}_{\lambda\rho}\frac{d x^\rho}{d\tau} V^\lambda~.
$$
Throughout this section, $\tau$ is an affine parameter along the worldline of the curve, i.e. $g_{\mu\nu}{dx^\mu\over d\tau}  {dx^\nu\over d\tau}  =1$. }
\be\label{eetmg}
S_{\rm EE}^{(\rm TMG)}= \int d\tau \bigg(m \sqrt{g_{\mu\nu}v^\mu v^\nu}   + s\, \tilde{n} \cdot \nabla n \bigg) ~,
\ee
where $m$ and $s$ are the mass and spin of the test particle, and $v^\mu\equiv {dx^\mu\over d\tau }$. As shown in \cite{Castro:2014tta}, the replica trick fixes these coefficients to be 
\be\label{eq:ms}
m={1\over 4G_3}~,\quad s={1\over 4G_3\mu}~.
\ee
The precession of the spinning particle is described by two normal vectors to the worldline of the particle: $n^\mu$ and $\tilde n^\mu$. The minimization of \eqref{eetmg} is subject to the constraints that $n^\mu$, $\tilde n^\mu$ and $v^\mu$  are mutually orthonormal:
\be
S_{\rm constraints}=\int d\tau \left[\lambda_1 n \cdot \tilde{n} + \lambda_2 n \cdot v +\lambda_3 \tilde{n} \cdot v +\lambda_4 (n^2+1) +\lambda_5(\tilde n^2-1)\right]~,
\ee
where $\lambda_i(\tau)$ are Lagrange multipliers. Note that $n^\mu$ is timelike, and $\tilde n^\mu$ is a spacelike vector. The resulting dynamics of the probe is given by 
\be\label{eq:mpd}
\nabla[m v^\mu + v_\rho \nabla s^{\mu\rho}] = -\frac{1}{2}v^\nu s^{\rho\sigma}R^\mu_{\,\,\,\nu\rho\sigma}~, \quad  s^{\mu\nu}  = - s \epsilon^{\mu\nu\lambda}v_\lambda~,
\ee
which is known as the Mathisson-Papapetrou-Dixon (MPD) equation, and the constraint
\be\label{eq:affine}
g_{\mu\nu}v^\mu v^\nu=1~.
\ee

We are interested in solving \eqref{eq:mpd}, and evaluating \eqref{eetmg}, for backgrounds of the form \eqref{eq:geng}. The first simplification in this case comes about by noting that the system preserves translation symmetries along $x^i=(x,t)$. This implies that \eqref{eq:mpd} contains two conserved quantities -- the momentum along $x$ and $t$-- which are
\bea\label{eq:pxt}
 P_x &=& m\, e^{2A} v^x + s e^{3A} (v^u)^2\partial_\tau\le(  {v^t\over v^u} \right) ~,\cr
P_t &=& m\, e^{2A} v^t + s e^{3A} (v^u)^2\partial_\tau\le(   {v^x\over v^u} \right) ~,
 \eea
with $P_x$ and $P_t$ the constants of motion. It is also useful to note that \eqref{eq:pxt}, together with \eqref{eq:affine}, implies 
\be\label{eq:vyp}
s \partial_\tau(e^A v^u) =P_x v^t-P_t v^x ~.
\ee
 The remaining component of the MPD equation \eqref{eq:mpd} is  the radial direction which reads
\be\label{eq:rad}
m\left[\partial_\tau(e^A v^u)+  e^{A}((v^t)^2 -(v^x)^2)\partial_uA \right]-s\, e^{-A} \partial_\tau\left(e^{3A}(v^t \partial_\tau{v}^x -v^x \partial_\tau{v}^t)\right) =0~.
\ee
Note that this equation is not independent: by taking derivatives of \eqref{eq:pxt} combined with  \eqref{eq:affine} gives \eqref{eq:rad}. 

%\footnote{ As it stands, the MPD equations are not only non-linear but as well heavily coupled. However, using \eqref{eq:pxt}-\eqref{eq:vyp}, we find the following equation for the radial component
%\be\label{eq:onlyvy}
%v^y \partial_\tau\le({1\over (v^y)^2} \le[m^2  - P_+P_-  e^{-2A}\ri]\ri)= -2s^2 e^{-A}\partial_\tau \le( v^y \partial_\tau \le(e^{A} {\dot v^y\over v^y}\ri)\ri) 
%\ee
%In the following sections we will not make use of \eqref{eq:onlyvy} but it might be useful for the purpose of finding exact solutions to the system.
%}

\subsubsection{No static solutions} 
Before discussing the solutions to \eqref{eq:pxt}-\eqref{eq:rad} that will be anchored at the boundary, it is instructive to highlight a general property of the MPD equations. Consider taking a static ansatz, i.e. 
\be\label{eq:smpd1}
v^t={dt\over d\tau}=0 \quad \Rightarrow \quad (v^u)^2 + (v^x)^2=e^{-2A}~.
\ee
In this case the MPD equations \eqref{eq:pxt} reduce to
\bea\label{eq:al}
 P_x = m\, e^{2A(u)} v^x  ~,\quad 
P_t=-s\, e^{A(u)}{v}^x \partial_uA(u) ~,
\eea
where we used \eqref{eq:smpd1}.  Assuming that both $s\neq 0$ and $m\neq0$, it is clear that \eqref{eq:al} is only consistent with  both $P_x$ and $P_t$ being constant if $e^{A(u)}=u^{-1}$. This choice of $A(u)$ corresponds to having exactly AdS$_3$ as a background and, moreover, $v^\mu$ would follow the usual geodesic path. The other obvious possibility is to take $v^x=0$ in \eqref{eq:al} but this would correspond to a solution that is disconnected --the endpoints of a single curve are not anchored at the UV boundary.

Hence, for a general background with $e^{A(u)}\neq u^{-1}$ and $(m,s)$ non-zero, the choice $v^t =0$ will not allow connected solutions to the equations of motion of the spinning particle. One might worry about this issue: it has been argued extensively in the literature that for a static background the Ryu-Takayanagi curve must lie on a constant time slice \cite{Headrick:2007km,Hubeny:2007xt}. The MPD equations clearly violate this restriction. Even though the background \eqref{eq:geng} is static, the gravitational Chern-Simons term is not invariant under $t\to -t$. We suspect that due to the lack of parity invariance of the theory there is no good reason to restrict the curve to lay on a space-like slice. 

Finally, we note that for $e^{A(u)}= u^{-1}$ and $v^t=0$, the conserved quantities satisfy
\be
m P_t= s P_x~. 
\ee

\subsubsection{Connected solutions: perturbative analysis}

We have established that connected solutions to the MPD equations must have $v^t\neq 0$ for non-AdS backgrounds, which makes the task of building connected solutions much harder.  We will first build these solutions by perturbing the MPD equations in $s$ (the spin of the probe). Our goal in the following will be to find a relation between the $z_0$ (the deepest point of the path) and the size $R$ of the boundary region; the aim is to show a breakdown of the relations \eqref{eq:hlmm}-\eqref{eq:hlmm1}.

To solve perturbatively the MPD equations it is useful to use the radial coordinate as the  proper length of the curve. In this gauge we have
\be\label{eq:radialu}
\partial_\tau = v^u \partial_u ~, \quad  \hat v(u)\equiv e^Av^u = (1 -(\partial_u t)^2+ (\partial_u x)^2)^{-1/2}~.
\ee
Using this parametrization \eqref{eq:pxt} becomes
\bea\label{eq:radg}
P_x &=& m \,e^{A}\hat v\, \partial_u x + s \,\hat v^3 \partial_u^2 t ~,\cr
P_t &=& m \,e^{A}\hat v\, \partial_u t + s \,\hat v^3 \partial_u^2 x ~.
\eea
 %
%We first look for solutions to \eqref{eq:radg} perturbatively in the spin of the probe particle $s$. In particular, 
%
We are interested in solutions to the MPD equations that are anchored at the boundary, i.e. the boundary conditions we will impose are 
\be\label{eq:bcmpd}
R\equiv \Delta x = x(u_f)- x(u_i)~,  ~\quad  \Delta t = t(u_f)- t(u_i) = 0~, \quad u_i, u_f\to 0 ~.
\ee
Taking $s$ as a small parameter, we expand the curve as
\bea\label{eq:xts}
x(u)= x_0(u) + \sum_{i=1}^\infty s^i x_i(u)~,\quad
t(u)= t_0(u) + \sum_{i=1}^\infty s^i t_i(u)~,
\eea
and for the conserved charges
\bea\label{eq:ptxs}
P_{t,x}= p^{t,x}_0 +\sum_{i=1}^\infty s^i  p^{t,x}_i~.
\eea
Using \eqref{eq:xts} and \eqref{eq:ptxs}, we will solve the MPD equations \eqref{eq:radg} order by order in $s$. Note that the spin $s$ is not quantized since the probe is a classical particle. From \eqref{eq:ms}, small $s$ and $m$ fixed is equivalent to taking $\mu$ large and $G_3$ fixed. 

We start at order $s^0$: the solutions are just those satisfying the geodesic equation. In accordance to the boundary conditions \eqref{eq:bcmpd}, at this order we will use the static geodesic:
\be
p_0^t =0 ~, \quad t_0(u)=0~,
\ee
and 
\be\label{eq:x0}
\partial_u x_0 = {z(u) \over \sqrt{z_0^2 -z(u)^2}}  ~,\quad z_0\equiv m/p_0^x  ~.
\ee
Looking ahead, we re-introduced $z(u)\equiv e^{-A(u)}$, which brings the line element \eqref{eq:geng} to the form \eqref{eq:zg}, and $z_0$ which will be turn out to be the deepest point of the curve. In this notation $\partial_u z= f^{1/2}(z)$ with $f(z)$ defined in \eqref{eq:zg}.
%For easy  comparison with \eqref{eq:zg}, we will also use
%\be
%\partial_y A(y) = f(z)^{1/2}~,\quad \partial_y = -{1\over z f^{1/2}} \partial_z~.
%\ee

At linear order in $s$, the first equation in \eqref{eq:radg} receives no correction due to the spin coupling --i.e. the last term in \eqref{eq:radg}. For this reason we set 
\be\label{eq:x1}
p_1^x=0~, \quad  \partial_u x_1 = 0 ~, 
\ee 
and hence at linear order in $s$ we find no correction to $R$. The second equation in \eqref{eq:radg} gives at linear order
\bea\label{eq:t1}
\partial_u t_1 &=& {1\over p_0^x}\left(p_1^t \partial_u x_0 +\partial_u \hat v_0\right) \cr
&=& {z\over p_0^x\sqrt{z_0^2 -z^2}}\left(p_1^t- {f(z)^{1/2}\over z_0} \right)~.
\eea
From here it is clear that the solution is not static: we cannot set $t_1$ to be a constant without drastic consequences. Imposing \eqref{eq:bcmpd}, which sets $\Delta t_1=0$, gives%\footnote{Note that $\hat v_0$ changes signs as it crosses the turning point. Hence, when integrating $\hat v_0$ to obtain \eqref{eq:p1fix} it is convenient to define $z=z_0 \cos(\zeta)$ and perform the integral with respect to $\zeta\in[-\varepsilon, \varepsilon]$. Here $\varepsilon\to \pi/2$ controls the endpoint position and UV cutoff of the boundary theory. For AdS$_3$ we have $p_1^t=1/z_0$.}
\be\label{eq:p1fix}
p_1^t = {2\over \Delta x_0}  ~,
\ee
where we used that $\Delta \hat v_0= \hat v_0 (u_f) -  \hat v_0 (u_i)=2$; $\Delta x_0$ is the length of the boundary interval  in the geodesic limit. Note that \eqref{eq:p1fix}, and in general $t(u)$, can be easily obtained from \eqref{eq:vyp}. For our boundary conditions \eqref{eq:bcmpd} the exact relation is 
\be
P_t = {s\Delta \hat v \over R}~.
\ee

Moving on to second order in $s$, the MPD equations give for the temporal component
\be
\partial_u t_2  = {p_2^t\over p_0^x}\partial_u x_0~,
\ee
where we made use of  \eqref{eq:x1}. Since we require $\Delta t=0$ this sets $p_2^t=0$, and hence without loss of generality we set $t_2=0$.  For the spatial component, the quadratic correction in $s$  is 
\bea\label{eq:x2e}
{m\over z(u)} \partial_u x_2&=& {p_2^x\over (\hat v_0)^{3}}  -\partial_u^2 t_1-{p_0^x\over 2\hat v_0} (\partial_ut_1)^2  \cr
&=&{1\over 2 p_0^x \hat v_0}\left[{z\over z_0^2}\partial_zf+\le(p_1^t -{f^{1/2}\over z_0}\ri)^2\right] + {1\over  (\hat v_0)^3}\left[{1\over 2p_0^x}\le({f\over z_0^2}-(p_1^t)^2\ri) +p_2^x\right]~.
\eea
In the first line we already made use of \eqref{eq:x1}, and from the first to the second equality we used \eqref{eq:t1}. The value of $p_2^x$ is
\be
p_2^x = {1\over 2p_0^x}\le({(p_1^t)^2}-{f(z_0)\over z_0^2}\ri)~,
\ee
which is determined by demanding convergence of $x_2$ when integrating \eqref{eq:x2e}. Replacing $p^x_2$ gives
\bea
\partial_z x_2={z\over 2 m^2 z_0 \hat v_0}\left[ z\partial_zf(z) + (z_0p_1^t -f(z)^{1/2})^2 +{1\over (\hat v_0)^2}(f(z)-f(z_0))\right] {1\over  f(z)^{1/2}} ~.
\eea
From here the quadratic correction to $\Delta x$ is
\bea
\Delta x_2 
=     {z_0\over m^2 }\int_{0}^{1} {dq\over  f(z_0 q)^{1/2}} {q\over (1-q^2)^{1/2}}  \left[ q\partial_qf(z_0q) + (z_0p_1^t -f(z_0 q)^{1/2})^2 +{(f(z_0q)-f(z_0))\over (1-{q^2})}\right]
\eea
where we defined $q=z/z_0$. In the small $z_0$ limit, which means that we are near the AdS boundary, we have $f(z_0q)\sim 1$ and the interval size is not corrected: $\Delta x_2 \sim 0$. This  simply implies that near the boundary the curve is well approximated by a geodesic. 

In the opposite regime, i.e. large $z_0$ limit, we have  $f(z_0q)\sim f_0(z_0 q)^n$ and  the interval is corrected as\footnote{$\alpha$ and $\beta$ are $\mathcal{O}(1)$ numbers coming from the integral over $q$. These numbers are not important for the general scaling with $z_0$ and are sensitive to the approximation chosen for the integral so we will not specify them further.}
 \bea\label{eq:dx2}
 R &=& \Delta x_0 + \Delta x_2   s^2+ O(s^4)\cr
 &\sim& 2\le( {\alpha z_0\over \sqrt{f_0 z_0^n}}+ O((f_0z_0^{n})^{-1/2})\ri) + {s^2\over m^2} \le(\beta z_0\sqrt{f_0 z_0^n }+ O((f_0z_0^{n})^{1/2})\ri)\,  + O(s^4)~,
 \eea
where for each order in $s$ we are writing the leading correction in $z_0$. These corrections are our first indication that there is a dramatic effect as we turn on the spin: the dependence of the interval size with $z_0$ is dramatically changed since now for $n\geq 2$ there is no bound on $R$ when $z_0$ is large. 

  As we go higher in perturbation theory it is simple to see that the structure of the corrections are even in $s$ for $x(u)$ and odd for $t(u)$. Moreover, a simple analysis of the structure of these corrections will imply that as we explore the limit $z_0$ large we will find
\be \label{subleading}
 \Delta x_{2i} \sim s^{2i} z_0^{1+(i-1)n/2}~.
 \ee 
 One might worry about the convergence of these corrections for arbitrarily large $z_0$ and  $s\lesssim 1$, and this is  legitimate concern. Our perturbative expansion in $s$ breaks down in the IR geometry, since we are neglecting derivatives in $z$ that become more important as $z_0$ is larger. The formulas above just illustrate that we can construct connected solutions, and that $R$ receives a non-trivial correction. In the following we will investigate the IR behavior of the solutions.

 \subsubsection{Connected solutions: IR behavior \label{IRbehaviour}}

The IR corrections we found  in the regime $s\ll m$ come with generically large coefficients as it is reflected in \eqref{eq:dx2}. For the purpose of convergence of this series, we will discuss the opposite regime: $s\gg m$.

Taking $m=0$ reduces the MPD equations \eqref{eq:radg} to 
 \bea\label{eq:gcg}
P_x =  s \,\hat v^3 \partial_u^2 t ~,\quad
P_t =  s \,\hat v^3 \partial_u^2 x ~.
\eea
 Two remarkable things happen in this limit. First, the equations do not depend on the warping factor of the metric: $e^{A(u)}$. Hence any solution we find in this regime is insensitive to details of the interior geometry. Second, in this regime me can solve the equations exactly. And related to this, in this limit we can recover static solutions: it is perfectly consistent to set $t(u)=0$ which just sets $P_x=0$ which is not possible if both $m$ and $s$ are non-zero.  
 
 On top of all of these elegant simplifications, there is a more interesting reason why to consider the limit $s\gg m$. In section \ref{sec:pir} we argued that the IR limit of the linearized equations suppressed terms like $e^A\partial_u$ versus $\partial_u^2$. Looking at \eqref{eq:radg} this would imply that for curves that explore the deep infrared geometry, the spin coupling terms dominates over the mass coupling. The limiting case in \eqref{eq:gcg} hence governs the IR dynamics, which as expected agrees that conformal gravity governs the long distance dynamics of the system.  
 
 In order to find solutions in this regime, it is convenient to rearrange \eqref{eq:gcg} as
 \be\label{eq:p321}
 P_t \,\partial^2_u t = P_x\,\partial^2_u x ~,\quad  s^2 \hat v^3 \partial_u^2 \hat v = P_x^2-P_t^2~.
 \ee
 The solutions to \eqref{eq:p321} are rather simple to obtain in general form. However, we are interested in solutions satisfying \eqref{eq:bcmpd}. This sets $P_x=0$ and $\partial_u t=0$. For the spatial direction we find
 %\bea
 %\partial_u x= a_0 \pm {a_1^3 h(u) \over \sqrt{s^2 - a_1^4 a_2 h(u)^2}}~,  \quad  \partial_u t= b_0 + {P_x\over P_t} \partial_u x~,
 %\eea
 %where 
 %\be
 %b_0 ={P_t^2-P_x^2\over P_tP_x} a_0~, \quad  a_1^2 = 1-{P_t^2-P_x^2\over P_x^2} a_0^2~, \quad a_2 =1-{P_x^2\over P_t^2} ~, 
 %\ee
 %and 
 %\be
 %h(u) = P_t u + b_1~.
 %\ee
 %Here $a_i$ and $b_i$ are constants. 
 \be
 \partial_u x= { h(u) \over \sqrt{ s^2 - h(u)^2}} ~,\quad h(u) = P_t u + h_1~.
 \ee
 with $h_1$ a constant. The turning point of this curve is defined by $h(u_0)= s$; $u_0$ is the analogous of $z_0$ used in the previous subsection. For this solution we find that
 \be
 R= 2\int_0^{u_0} \partial_u x = {s\over P_t} = u_0  + {h_1\over P_t}~.
 \ee
 The width of the spatial interval is finite and obviously independent of the details of the geometry in the interior, i.e. there is no appearance of $A(u)$. This should be compared with the geodesic solution using this parameterization:  $u_0$ is defined by $e^{-A(u_0)}= m/P_x$ (see \eqref{eq:x0}), which makes the dependence on the warping factor explicit.

The next step would be to match the solutions for small and large $s$. It is important to emphasize that the small $s$ expansion assumes that the subleading corrections \eqref{subleading} are small which implies that $\Delta x_0$ (or alternatively $z_0$) cannot be arbitrarily larger than the the mass gap (controlled by $f_0$). For $z_0$ arbitrarly large the perturbation in $s$ breaks down. Similarly, for large $s$ we are neglecting terms in the MPD equation that can become dominant for short intervals. Still, there will be a matching region where $\Delta x_0$ is of the order of the mass gap, and numerically one should be able find overlap of both regimes in $s$. As the matching is done, the integration constants (such as $h_1$) in the IR will be fixed by requiring continuity with the UV.   We will leave this for future work. 
 
\subsubsection{On-shell action of the spinning probe}\label{sec:eep}

We have shown, at least perturbatively in $s$, that we can design connected solutions to the MPD equations for arbitrary $R$. In the following we will quantify how these solutions modify the behaviour of entanglement entropy in the presence of the gravitational anomaly.

We will evaluate \eqref{eetmg} up to order $s^2$ for a connected solution. In order to do so, we need to build the normal vectors to our curve and impose appropriate boundary conditions. Recall that the normal vectors satisfy the following orthonormality constraints:
\be
S_{\rm constraints}=\int d\tau \left[\lambda_1 n \cdot \tilde{n} + \lambda_2 n \cdot v +\lambda_3 \tilde{n} \cdot v +\lambda_4 (n^2+1) +\lambda_5(\tilde n^2-1)\right] \,,
\ee
and that the tangent vector satisfies \eqref{eq:affine}.
%\be
%v^\mu = {d x^\mu \over d\tau} ~,\quad g_{\mu\nu}v^\mu v^\nu=1 \,.
%\ee
%For the metric given in \eqref{eq:uu}, this gives
%\be
%v^\mu=(v^u,v^t,v^x)~,\quad e^{2A}\le((v^u)^2-(v^t)^2+(v^x)^2\ri)=1
%\ee

It is rather simple to build normal vectors to $v^\mu=(v^u,v^t,v^x)$ for the background \eqref{eq:geng}. For instance,  two mutually orthonormal vectors are\footnote{Our convention for the epsilon tensor is
\be
\epsilon_{txu}=\sqrt{-g}~,\quad \epsilon^{txu}=-{1\over \sqrt{-g}}~.  \notag
\ee}
\be\label{eq:QQ}
Q^\mu = F(\tau)(0,v^x, v^t) \ \ \ \ \ \ \ \ \ \ \tilde Q^\mu = \epsilon^{\mu\nu\lambda} v_\nu Q_\lambda = \le(-{e^{-A}\over F},\,  F\,\hat v\, v^t,\,  F \,\hat v\, v^x\ri) \,.
\ee
Here is $Q$ is timelike and $\tilde{Q}$ is spacelike;  $F^{-2}=e^{2A}\le((v^x)^2-(v^t)^2\ri)$ which adjusts the normalization appropriately.  The vectors $(Q^\mu,\tilde Q^\mu)$ are however not very convenient to evaluate \eqref{eetmg} since they don't a priori  satisfy any reasonable boundary conditions. Following  \cite{Castro:2014tta}, it is natural to require that the timelike normal vector should point along the CFT time direction at the endpoints of the curve. This allows us to compare the orientation of normal vector at each endpoint as we transport them along the curve. 

To implement unambiguously any precise choice of boundary conditions, it is useful to construct normal vectors that are Fermi-Walker transported along the solution. More concretely, we want to first consider vectors $(q^\mu,\tilde q^\mu)$ that are  orthonormal  to $v^\mu$ and also satisfy
\bea\label{eq:11}
\nabla q^\mu = - (q_\alpha \nabla v^\alpha) v^\mu~,\quad \nabla \tilde q^\mu = - (\tilde q_\alpha \nabla v^\alpha) v^\mu~.
\eea
The advantage of this parametrization is that we can easily evaluate \eqref{eetmg} for our choice of boundary conditions. Using $(q^\mu, \tilde q^\mu)$, it is convenient to write  
\be
n^\mu = \cosh(\eta(\tau)) q^\mu + \sinh(\eta(\tau)) \tilde{q}^\mu ~,\quad  \tilde n^\mu = \epsilon^{\mu\nu\lambda} v_\nu n_\lambda~,  \label{nexp}
\ee
where $\eta(\tau)$ is a function that is adjusted such that at the endpoints\footnote{Here we define $\le(\p_t\ri)_{\rm CFT}$ as a vector that points in the time direction at the boundary, however it is normalized to ensure that $n^2=-1$ with respect to the bulk metric.}
\be
n_i = n_f = \le(\p_t\ri)_{\rm CFT}~. \label{CFTbc}
\ee
Given \eqref{nexp} and \eqref{eq:11}, the anomalous portion of \eqref{eetmg} reduces to 
\bea\label{eq:anomS}
S_{\rm anom} &=& s\, \int d\tau \, \tilde{n} \cdot \nabla n  \cr
 &=& s \le(\eta(\tau_f) - \eta(\tau_i)\ri) \cr
 &=& s \log\le(\frac{q(\tau_f) \cdot n_f - \tilde q(\tau_f) \cdot n_f}{q(\tau_i) \cdot n_i - \tilde q(\tau_i) \cdot n_i}\ri)~.
\eea
The appeal of this equation is that we actually don't need to solve for $\eta(\tau)$. We just need to build $(q^\mu, \tilde q^\mu)$ and evaluate the inner product with \eqref{CFTbc}.

To build $(q^\mu, \tilde q^\mu)$ it is actually useful to use  \eqref{eq:QQ}; in particular, it is rather clear that each basis of normal vectors should be related as follows 
\bea\label{eq:qh}
q^\mu &=& \cosh(h(\tau)) Q^\mu+ \sinh(h(\tau)) \tilde Q^\mu~,\cr
\tilde q^\mu &=& \sinh(h(\tau)) Q^\mu+ \cosh(h(\tau)) \tilde Q^\mu~.
\eea
The advantage is that we reduced the undetermined number of variables, since we only need to solve for $h(\tau)$ such that \eqref{eq:11} is satisfied. And this is a straight forward procedure: plugging  \eqref{eq:qh} in \eqref{eq:11}, the resulting equation for $h(\tau)$ is 
\bea \label{eq:dh}
\partial_\tau h(\tau) &=&e^{3A} F^2 v^u\le(v^t \dot v^x -v^x\dot v^t\ri)\cr
&=& {F^2\over s}\left(P_t v^t - P_x v^x+\frac{m}{F^2}\right) ~.
\eea
To obtain the first line, we just made use of \eqref{eq:QQ}; to obtain the second line we used the MPD equations \eqref{eq:pxt}. From here, we find that \eqref{eq:anomS} becomes
 \bea
S_{\rm anom} &=& s\log \le({( v^x -  \hat v\, v^t )_f \over  ( v^x -  \hat v\, v^t )_i}\ri)-s \int_{\tau_i}^{\tau_f} d\tau \partial_\tau  h(\tau)~.
\eea

With these results we can now evaluate \eqref{eetmg}. Including all contributions we find
\bea\label{eq:seexc}
S_{\rm EE}^{(\rm TMG)}&=& m\int d\tau  \sqrt{g_{\mu\nu}v^\mu v^\nu}   +  S_{\rm anom}\cr
&=& 2m\int_{\epsilon}^{z_0}\frac{dz}{z f(z)^{1/2}}\frac{1}{\hat{v}} +  s\log \le({( v^x -  \hat v\, v^t)_f \over  ( v^x -  \hat v\, v^t )_i}\ri)- 2s \int_{0}^{z_0} {dz\over f(z)^{1/2}} \partial_z  h(z)~.
\eea
At this stage, it is important to emphasize that we have made no approximations so far in $S_{\rm EE}^{(\rm TMG)}$. The expression above is exact and correctly implements the boundary conditions \eqref{CFTbc}. Also note that in the first integral we explicitly include a $UV$ cutoff $\epsilon$ since the proper distance is divergent. However, the last integral in \eqref{eq:seexc} does not carry divergences as $z\to 0$;  this will be evident shortly. 

% We split the order $s^2$ correction to $S_{\rm EE}$ into three pieces \,,
%\be
%S_1=s(h(\tau_f)-h(\tau_i)) \ \ \ \ \ \ \ \  S_2=s \log \left( \frac{(Fv^x-Fv^u v^te^A)_i}{(Fv^x-Fv^uv^te^A)_f}\right) \ \ \ \ \ \ \ \ S_3=S_{\text{length}}^{(2)}
%\ee

We now proceed to evaluate \eqref{eq:seexc} for our perturbative solution. The first integral, which measures the proper length, gives
\bea
2m\int_{0}^{z_0}\frac{dz}{z f(z)^{1/2}}\frac{1}{\hat{v}} &=&2m \int_{\epsilon}^{z_0}\frac{dz}{f(z)^{1/2}} {z_0\over z} {1\over \sqrt{z_0^2-z^2}}  \cr &+&  \frac{ s^2}{m} \int_{0}^{z_0}\frac{dz}{f(z)^{1/2}}\frac{z(z_0^2(f(z)-f(z_0))+z(z_0^2-z^2)\partial_z f)}{z_0(z_0^2-z^2)^{3/2}} + O({s^4\over m^3}) \,.
\eea
And for the terms in $S_{\rm anom}$ we obtain
\bea
2s \int_{0}^{z_0} {dz\over f(z)^{1/2}} \partial_z  h(z)=\frac{s^2}{m}\int_{0}^{z_0}{dz\over f(z)^{1/2} } \sqrt{1-\frac{z^2}{z_0^2}} \partial_z f  + O({s^4\over m^3})~,
\eea
and
\be
s\log \le({( v^x -  \hat v\, v^t)_f \over  ( v^x -  \hat v\, v^t )_i}\ri)=-\frac{2s^2}{m}\left( \frac{2z_0}{\Delta x_0}-1\right)~.
\ee
Combining these three contributions gives
\bea
S_{\rm EE}^{(\rm TMG)}&=& 2m \int_{\epsilon}^{z_0}\frac{dz}{f(z)^{1/2}} {z_0\over z} {1\over \sqrt{z_0^2-z^2}}  -\frac{2s^2}{m}\left( \frac{2z_0}{\Delta x_0}-1\right)\cr &&+  \frac{ s^2}{m} \int_{0}^{z_0}\frac{dz}{f(z)^{1/2}} \le( \frac{ z z_0(f(z)-f(z_0))}{(z_0^2-z^2)^{3/2}}+\partial_z f \frac{2z^2-z_0^2}{z_0 \sqrt{z_0^2-z^2}} \ri)+ O({s^4\over m^3}) ~.
\eea
%A quick calculation gives
%\bea
%S_1&=&\frac{s^2}{m}\int_{0}^{z_0}dz\frac{\partial_z f }{f(z)^{1/2}}  \sqrt{1-\frac{z^2}{z_0^2}} \notag \\
%S_2&=&-\frac{2s^2}{m}\left( \frac{2z_0}{\Delta x_0}-1\right) \\
%S_3&=&\frac{ s^2}{m} \int_{0}^{z_0}\frac{dz}{f(z)^{1/2}}\frac{z(z_0^2(f(z)-f(z_0))+z(z_0^2-z^2)\partial_z f)}{z_0(z_0^2-z^2)^{3/2}} \notag
%\eea

In the limit $z_0\ll 1$ where $f(z)\sim 1$, one immediately finds $S_{\rm EE}^{(\rm TMG)}$ reduces to the geodesic length. This is of course expected as we are only probing the geometry near the boundary. The other interesting regime is for large $z_0$ where we use $f(z)\sim f_0 z^n$. %One finds that all three pieces have the same scaling with $z_0$, namely
%\bea
%S_{\rm EE}^{(2)} \sim \frac{z_0^{n/2}}{m} \,.
%\eea
%Combining with the order $s^0$ contribution we find
We find
\be
S_{\rm EE}^{(\rm TMG)}\sim \left(m z_0^{-n/2}+{s^2\over m} z_0^{n/2}+\ldots\right)~,
\ee
where ``$\ldots$'' represents corrections both in $s^2$ and in $1/z_0$. We immediately see that the order $s^2$ drastically changes the behaviour of the entanglement entropy. More importantly, by trading $z_0$ with $R$ via \eqref{eq:dx2},  $S_{\rm EE}^{(\rm TMG)}$ is always dependent on the size of the interval. This is generic for gapless systems as expected.

%To caracterize an RG flow, It is usually interesting to study the following quantity \cite{Liu:2013una} which has a natural c-function interpretation:
%\be
%\mathcal{R}=R \frac{d S_{\rm EE}}{dR} \,.
%\ee
%We can expand this quantity in powers of $s$ and we find
%\be
%\mathcal{R}\sim z_0^{-n/2}+s^2 z_0^{n/2}+...
%\ee

 %%%%%%%%%%%%%%%%%%%%%%%%%%%%%%%%%%%%%%%%%%%%%%%%%%%%%%%%%
%%%%%%%%%%%%%%%%%%%%%%%%%%%%%%%%%%%%%%%%%%%%%%%%%%%%%%%%%

\section{Discussion}

In this work we discussed  properties of RG backgrounds in three dimensional gravity in the presence of a gravitational anomaly.  Our main result was to show that the infrared geometry could never be interpreted as the dual of a gapped system, as expected from general QFT arguments. The geometry did not make evident that a gravitational anomaly is protected; the analysis presented here is an attempt to make these universal features robust and in accordance with the RG theorems in the dual theory.  

The most straightforward computation that captures how a chiral anomaly is protected is the spectrum of fluctuations.   We showed that, regardless of the radial profile of the background geometry, the linearized spectrum of gravitational perturbations has a continuous sector deep in the IR. This can be a rather cumbersome task; a simpler holographic observable is entanglement entropy which we discussed at length.

As expected, holographic entanglement entropy behaved in accordance with RG theorems: we showed that connected solutions are always present in any vacuum solution in the presence of the gravitational anomaly, which makes entanglement always sensitive to the size of the interval as we flow from the UV to the IR. This is strikingly different than in the absence of such anomaly. The dynamics of the probe that measures entanglement is described by the MPD equations rather than geodesics \cite{Castro:2014tta}, and our analysis shows that it exhibits the right features that explains why the system is not gapped in the IR. 

As highlighted in \cite{Castro:2014tta,Azeyanagi:2015uoa}, one of the interesting properties of the contribution of the gravitational anomaly to entanglement entropy is that the answer is now sensitive to the choice of Lorentz frame. To be more precise, if we rotate our normal vector by an $SO(1,1)$ transformation
\be
n^a ~\to ~\Lambda^a_b(\eta')\, n^b
\ee
 with boost parameter $\eta'$, then, due to \eqref{eq:anomS}, entanglement entropy transforms as
\bea
S_{\rm EE}^{(\rm TMG)} ~\to~ S_{\rm EE}^{(\rm TMG)} +  {c_L-c_R\over 12}\, \Delta \eta' ~.
\eea
This change is universal, and could be interpreted as one way to ``measure'' the anomaly from entanglement. This derivation holds both in the bulk and boundary theory, however, from the bulk point of view these manipulations are only consistent if there is a smooth connected worldline. For that reason it is crucial to explicitly construct connected solutions, which is a rather non-trivial task.

We were able to construct only perturbative solutions to the MPD equations in two different regimes: small and large $s$. It would be very interesting to  find exact solutions and hence to have a better understanding of $S_{\rm EE}^{(\rm TMG)}$ as a function of $s$. Another interesting feature that we found was that any connected solution does not lie on a constant time slice, despite the fact that the background is stationary. This should be contrasted to the original Ryu-Takayanagi prescription. It is not clear if this feature will persist in the higher dimensional examples discussed
in \cite{Guo:2015uqa,Azeyanagi:2015uoa,Nishioka,IqbalWall},  but it is worth further investigation.

Along the lines of the construction in \cite{Myers:2012ed,Liu:2013una}, it would be interesting to build a monotonic function that controls the RG flow out of holographic entanglement entropy. From our explicit results in section \ref{sec:eep}, it is not evident how to build such quantity from $S_{\rm EE}^{(\rm TMG)}$.\footnote{It is also the case that our bulk theories are not generically unitary, since TMG carries negative energy excitations. This feature makes it confusing whether we should even demand that any candidate $c$-function is monotonic.} However, a more interesting route is to modify the construction in \cite{Casini:2011kv}: for a QFT with a chiral anomaly, can we derive the constraints on $c_{L,R}$ along a RG flow via appropriate inequalities associated to entanglement entropy? We leave this question for future work.

\section*{Acknowledgements}

L.Y.H. would like to thank X-G Wen for first coming up with the question and for many inspiring discussions. We would like to specially thank Nabil Iqbal for many interesting and helpful discussions related to the work presented here. We also thank  Tatsuo Azeyanagi, Jan de Boer, Kristan Jensen, Rob Myers, Gim S. Ng, and Charlotte Sleight  for useful discussions. 
A.C. is supported by Nederlandse Organisatie voor Wetenschappelijk Onderzoek (NWO) via a Vidi grant. A.B. is supported by the Swiss National Science Foundation (SNF), Grant No. P2SKP2\textunderscore158696.

 %%%%%%%%%%%%%%%%%%%%%%%%%%%%%%%%%%%%%%%%%%%%%%%%%%%%%%%%%
%%%%%%%%%%%%%%%%%%%%%%%%%%%%%%%%%%%%%%%%%%%%%%%%%%%%%%%%%

\appendix

\section{Conventions}\label{app:cn}

Throughout this work we will use the following coordinate system for the RG backgrounds:
\bea
ds^2 &=& {dz^2\over z^2 f(z)} +\frac{\eta_{ij} dx^i dx^j}{z^2} \cr 
&&\cr
%&=& dy^2 + e^{2A(y)} \eta_{ij} dx^i dx^j\cr
%&&\cr
&=& e^{2A(u)}\le(du^2 +  \eta_{ij} dx^i dx^j\ri)~.
\eea
The relations between various definitions gives
\be
z=e^{-A}~,\quad~, \quad \partial_u =   f^{1/2} \partial_z~.
\ee

In the limit IR limit, for $0<n<4$, we have
\bea\label{eq:uu1}
&u\sim z^{-n/2+1} \to \infty ~\quad& {\rm if}~ 0<n<2~,\cr
&u\sim \log(z) \to \infty ~\quad& {\rm if}~ n=2~,\cr
&u\sim u_0+ z^{-n/2+1} \to u_0 ~\quad& {\rm if}~ 2<n<4~,
\eea
and in the UV limit we have
\be
z\to 0~, \quad f(z)\sim 1~, \quad  u\sim z~.
\ee

\section{Linearized analysis}\label{app:1}
In this appendix we derive the master equations for the metric and scalar fluctuations at linear level. 
The action is
\be
I[g,\phi] = \frac{1}{16\pi G_3} \int d^3x \sqrt{g} \left[R - 8( \nabla\phi)^2 - V(\phi)\right] + I_{\rm GCS}~,
\ee
and the equations of motion are
\bea\label{eq:EHS}
R_{\mu\nu} -{1\over 2} g_{\mu\nu} R+{1\over \mu} C_{\mu\nu}= 8 \partial_\mu \phi \partial_\nu \phi - 4 g_{\mu\nu} \partial^\alpha \phi \partial_\alpha \phi -{1\over 2} g_{\mu\nu} V(\phi)
\eea
and
\bea\label{eq:scalareom}
16\nabla^2 \phi - {\partial \over \partial \phi} V(\phi) &=&0~.
\eea
To start we will keep the potential $V(\phi)$ arbitrary, and it will be specified as needed.
%
%\be
%V(\phi)=-2 (\cosh^6\phi-\sinh^6\phi)=-{1\over 4} (5+3\cosh (4\phi))~.
%\ee

We setup the linearized analysis by first defining
\be\label{eq:g1}
g_{\mu\nu}= g^{(0)}_{\mu\nu} + h_{\mu\nu} ~,\quad \phi = \phi_{(0)} + \delta\phi~.
\ee
Here $(g^{(0)}_{\mu\nu}, \phi_{(0)})$ define the background solution, and $(h_{\mu\nu},\delta\phi)$ are the small perturbations around the background. We use a radial gauge $h_{u\mu}=0$. For the background we have 
\be\label{eq:bb11}
ds^2_{(0)} =g^{(0)}_{\mu\nu}dx^\mu dx^\nu= e^{2A(u)}\le(du^2+dw d\bar w\ri)~, %\quad \phi_{(0)}= \tanh^{-1}\left(\sqrt{B\,r(y)\over H^2}\right)~,
\ee
and we will use null coordinates for the boundary directions: 
\be
w=-t + x ~,\quad \bar w = t+x ~, \quad \partial \equiv\partial_w~,\quad \bar\partial\equiv \partial_{\bar w}~.
\ee

When manipulating the equations, the following identities are useful
\bea\label{eq:ap1}
8(\dot \phi_{(0)})^2 +\ddot A(u) - (\dot A(u))^2=0~,\quad  V_0 e^{2A}+2\ddot A+8\dot\phi_0^2=0
\eea
which are due to \eqref{eq:EHS} at zeroth order, and the scalar equation \eqref{eq:scalareom} gives
\bea\label{eq:ap2}
{1\over16} {\partial V_0 \over \partial \phi}  = e^{-2A}\le(\dot A(y) \dot \phi_{(0)}+ \ddot\phi_{(0)}\ri)~.
\eea
Here, and through out this appendix, prime denotes derivative with respect to $u$, and 
\be
V_0\equiv V(\phi_{(0)})~,\quad \dot\phi_0\equiv \partial_u\phi_0~, \quad \dot A \equiv \partial_u A(u)~.
\ee
%We will work in the radial gauge where
%
%\be\label{eq:h1}
%\hat h_{y\mu}=0~,
%\ee
%and due to the symmetries of the background we will decompose the fluctuations in Fourier modes:
%\be\label{eq:h2}
%\hat h_{ij} =  e^{i\omega t+ik x} e^{2A(y)} h_{ij}(y) ~,\quad \delta\phi =  e^{i\omega t+ik x} \delta\phi(y) 
%\ee

The linearized metric pertubations are constructed as follows. Plugging in \eqref{eq:g1} in \eqref{eq:EHS}, we get that the $(i,j)$ components of \eqref{eq:EHS} are:
\bea\label{eq:yi1}
{1\over \mu}\bigg(\partial_u^2 \bar\partial  h_{w w} -{2}\partial^2 \bar\partial h_{w\bar w} +\partial \bar\partial^2 h_{w w} +\partial^3  h_{\bar w\bar w} \bigg)+e^A \partial_u\partial  h_{w\bar w}
- e^A\partial_u \bar\partial h_{w w} +8e^A\dot\phi_{(0)}\,\partial \delta \phi=0~,\cr
{1\over \mu}\bigg(\partial_u^2\partial  h_{\bar w \bar w} -2\partial\bar\partial^2 h_{w\bar w} + \bar\partial^3 h_{w w} +\partial^2\bar\partial  h_{\bar w\bar w}\bigg) +e^A\partial_u\partial  h_{\bar w\bar w}
- e^A\partial_u\bar\partial  h_{w \bar w}-8e^A\dot\phi_{(0)}\,\bar\partial \delta \phi=0~,
\eea
and
\be\label{eq:trace1}
\partial_u\le( 8e^{A}\dot\phi_{(0)}\delta  \phi+ e^A\partial_u  h_{w\bar w}  -{1\over \mu}\big(\bar\partial^2 h_{ w w}-\partial^2 h_{\bar w\bar w}\big)\ri)=0~.
\ee

The trace of  \eqref{eq:EHS} gives
\be\label{eq:r2}
-\partial^2h_{\bar w\bar w}-\bar\partial^2h_{ww}+2\partial\bar\partial h_{w\bar w}+ e^{-2A}\partial_u\left(e^{2A}\partial_u{h}_{w\bar w} \right) +12(\ddot\phi_{(0)}+\dot A \dot\phi_{(0)})\delta \phi+4\dot\phi_{(0)}\partial_u\delta\phi=0~.
\ee
However, this is not an independent  equation: taking a derivative of \eqref{eq:r2}, and after using heavily \eqref{eq:yi1}-\eqref{eq:trace1}, one obtains
\be\label{eq:kgl1}
e^{-A}\partial_u( e^{A}\partial_u\delta \phi)+4\partial \bar \partial \delta\phi - e^{2A} {1\over 16} {\partial^2 V_0\over \partial \phi^2}   \delta\phi +2 \dot \phi_{(0)}  \,\partial_u h_{w\bar w}=0
\ee
which is the equation obtained by linearizing \eqref{eq:scalareom}. Hence the independent equations that describe the fluctuations are \eqref{eq:yi1}-\eqref{eq:trace1} and  \eqref{eq:kgl1}. Equation \eqref{eq:r2} serves as a constraint to fix integration constants.   In comparison with prior literature, our equations agree with  \cite{Skenderis:2009nt}, provided we set $\delta \phi=0$ and the background metric is AdS$_3$; note that $(\partial_\rho)_{\rm here}=-2(\partial_\rho)_{\rm there}$.

\subsection{Einstein gravity plus scalar system}\label{app:EES}

Setting $1/\mu$ to zero simplifies significantly the analysis.  The linearized equations \eqref{eq:yi1} and \eqref{eq:trace1} simplify to 
\bea\label{eq:ijeh}
&&e^A\partial_u \bar\partial h_{w w} =\partial K_1~,\cr
&&\cr
&&e^A\partial_u\partial  h_{\bar w\bar w}=\bar\partial K_1~,\cr
&&\cr
&& 8e^{A}\dot\phi_{(0)}\delta  \phi+ e^A\partial_u  h_{w\bar w} =K_1~.
\eea
where $K_1$ is an integration constant coming from integrating \eqref{eq:trace1}. \footnote{Note that we use the term integration constant loosely as $K_1$ is a function of $x$ and $t$. We will use the notation where $K_1$ is also expanded in Fourier modes in which case $\partial$ and $\bar\partial$ can be viewed merely as numbers. We will use this convention for the remainder of the section and apologize if it creates any confusion.} The components for $h_{ww}$ and $ h_{\bar w \bar w}$ of the metric are independent of $\delta \phi$ and can be solved easily provided a warping factor  $e^{2A(y)}$ is specified. 
Using \eqref{eq:ijeh}, the scalar equation \eqref{eq:kgl} reduces to
\be\label{eq:kgl}
\partial_u^2\psi+(\omega^2-k^2)\psi + W_{\rm eff}(u) \psi +2K_1e^{-A/2}\dot\phi_{(0)}=0~.
\ee
where $\psi\equiv e^{A/2}\delta \phi$. This is a Schrodinger equation (with a inhomogeneous term proportional to $K_1$) and effective potential
\bea
W_{\rm eff}&=& -  e^{2A}{1\over 16} {\partial^2 V_0\over \partial \phi^2}-16 (\dot\phi_{(0)})^2 +e^{-A/2}\partial_u(e^{A}\partial_u e^{-A/2})\cr
&=&-  e^{2A}{1\over 16} {\partial^2 V_0\over \partial \phi^2}-{9\over 4}(\dot A)^2+ {3\over 2}\ddot A
\eea
In the IR, the effective potential behaves as
\be
W_{\rm eff}\sim \frac{1}{(u-u_0)^2}\sim z^{n-2}
\ee
The homogeneous part of the solution for $\delta\phi$ is thus similar to that of a probe field in the IR. From this we conclude that the system is gapped if $n>2$.

\subsection{Decoupling of the equations}\label{app:deq}

Here we briefly describe how to decouple the linearized equations for the metric perturbations: we will find a set of differential equations that only involve $(h_{w\bar w},\delta\phi)$.

 The key is to take two derivatives of the constraint \eqref{eq:r2} and then use repeatedly \eqref{eq:yi}, \eqref{eq:trace} and \eqref{eq:r2} to eliminate $(h_{\bar w\bar w},h_{ww})$. The resulting equation is
\bea\label{eq:mastereq}
&&\le[ \partial_\rho( e^{-4A} \partial_\rho) + e^{-2A}4\partial \bar \partial -\left(\mu^2-e^{-4A} A'^2\right)\ri]  h''_{w\bar w}-{e^{-2A}\over 2}\partial_\rho^3(e^{-2A}) h'_{w\bar w} \cr
&&+e^{-2A} X'' - e^{-4A} A' Y'' +2 \partial \bar \partial X -2A' e^{-2A} \partial \bar \partial Y -\left(\mu^2-e^{-4A} A'^2\right) Y'=0
\eea
where 
\be
X\equiv 4 e^{-2A}(3 \phi''_{(0)}\delta \phi + \phi'_{(0)}\delta \phi' )~,\quad Y\equiv 8\phi'_{(0)}\delta \phi ~,
\ee
and for compactness we introduced the radial variable $\rho$ which is given by
\be
X'\equiv \partial_\rho X \equiv -e^{A}\partial_u X~.
\ee
In this  notation the linearized Klein-Gordon equation \eqref{eq:kgl1} reads
\be\label{eq:newkg1}
e^{-4A}\delta \phi''+e^{-2A}4\partial \bar \partial \delta\phi -  {1\over 16} {\partial^2 V_0\over \partial \phi^2}   \delta\phi +2 e^{-4A}\phi'_{(0)}  \, h'_{w\bar w}=0~.
\ee

Replacing \eqref{eq:newkg1} into \eqref{eq:mastereq}, we get a fith order differential equation for $\delta\phi$. This equation is difficult to solve but one can study its IR limit. Taking this limit induces one simplification in the equation: the $\mu^2$ terms are subleading as $\mu^2\ll e^{-4A} A'^2$ in the IR. The same mechanism responsible for the decoupling of the $h_{ww}$ $h_{\bar w\bar w}$ is present here, the gravitational Chern-Simons term is dominating in the IR. However even upon taking an IR limit we cannot solve this equation for general $n$, but we can get an analytic solution for the special case $n=2$. The solutions behave as
\be
\delta\phi \sim z^{\sqrt{\frac{5}{4}-\partial\bar\partial}} \,.
\ee
This behavior is essentially similar to that of the $n=2$ case for a probe field \cite{Liu:2013una}. It is thus likely that the $h_{w\bar w}$ and $\delta\phi$ fields have the same properties as a massless probe field and are gapped for $n>2$. However, they only act as a source term for the diagonal components of the metric whose homogeneous equations still lead to ungapped modes.

\section{$U(1)$ gauge anomalies}\label{sec:u1}

Another very interesting example where the boundary CFT can be made chiral and therefore stable against forming a mass gap is a bulk theory with the addition of gauge Chern-Simons terms.  For simplicity, let us consider $U(1)$ CS theories, contributing to the action the following terms
\be\label{eq:csact}
S_{CS} = \frac{k_L}{4\pi} \int A\wedge d A  - \frac{k_R}{4\pi} \int \tilde{A}\wedge d\tilde{A}~,
\ee
where $k_L$ and $k_R$ are positive and $A$ and $\tilde{A}$ will lead to a left and right moving respectively $U(1)$ current algebra at level $k_L$ and $k_R$ at the boundary \cite{Kraus:2006wn,Jensen:2010em,Keranen:2014ava}. What is interesting about this scenario is that, as for the example with gravitational anomalies in section \ref{sec:tmg}, the Chern-Simons action \eqref{eq:csact} does not affect the background \eqref{eq:zg}. Therefore, in a holographic setup there is a potential ambiguity if the system is truly gapped or not.

Before considering what is the effect of \eqref{eq:csact} in the bulk, we will review the CFT$_2$ reasoning behind the fact that  the chiral anomaly is protected under RG. For any Lorentz invariant theory, $k_L- k_R$ is conserved along the RG flow of any $1+1$ dimensional theories that is a CFT in the UV limit. One version of this proof is as follows: current conservation dictates
\be\label{eq:cf}
\partial_\mu j^\mu = \frac{(k_L-k_R)}{2\pi} \epsilon^{\mu\nu}F_{\mu\nu}~,
\ee
where the right hand side arises from quantum anomaly and $F$ is a background gauge field  that is coupled to the global symmetry.
Due to this anomaly, the two point functions are 
\bea\label{eq:cc1}
\langle j_w(w,\bar{w}) j_w(0) \rangle &=& \frac{k_L(|w|)}{w^2} \notag \\
\quad \langle j_w(w,\bar{w}) j_{\bar{w}}(0)\rangle &=& \frac{a(|w|) \eta_{w\bar{w}} + b(|w|) w \bar{w}}{|w|^2} \\
\quad \langle j_{\bar{w}}(w,\bar{w}) j_{\bar{w}}(0) \rangle &=& \frac{k_R(|w|)}{\bar{w}^2}~, \notag
\eea
where $w= x+i t$ and $\bar{w} = x-i t$. The above relations follow only from Lorentz invariance, which leaves at this stage the functions $a(|w|)$,  $b(|w|)$, $k_{L,R}(|w|)$ arbitrary. Our goal in the following will be to restrict their dependence on $|w|$. 

We first note that for a CFT in the UV, $k_L, k_R$ approaches a constant as $w\to 0$ and that $\langle j_w j_{\bar{w}}\rangle \to 0$. 
Note also that since $j_w$ and $j_{\bar{w}}$ commute when they are not coincident, it means that 
\be\label{eq:cc2}
\langle j_w(w,\bar{w}) j_{\bar{w}}(0)\rangle  = \langle j_{\bar{w}}(w,\bar{w}) j_{w}(0)\rangle ~.
\ee

Now we can use current conservation to relate the correlators given above. We are not coupling the global symmetry to any external field, so current is in fact conserved. Moreover, even if there is a non-trivial background field, the right hand side of the conservation equation \eqref{eq:cf} only leads to extra contact terms that would not contribute when the currents are not inserted at coincident points. 
Therefore we have 
\be\label{eq:cc3}
\langle \bar{\partial} j_w(w,\bar{w}) j_w(0)\rangle = -\langle \partial j_{\bar{w}}(w,\bar{w}) j_w(0)\rangle~,
\ee
 leading to
\be\label{dkL}
\frac{\bar{\partial} k_L(|w|)}{w^2} = \bar{\partial} (\frac{a(|w|) \eta_{w\bar{w}} + b(|w|) w \bar{w}}{w^2})~,
\ee
where we used \eqref{eq:cc1} and \eqref{eq:cc2}. One can repeat the same exercise with $\langle j_{\bar{w}}(w,\bar w) j_{\bar{w}(0)}\rangle$ replaced appropriately in \eqref{eq:cc3} and obtain
\be \label{dkR}
\frac{\bar{\partial} k_R(|w|)}{\bar{w}^2} = \bar{\partial}(\frac{a(|w|) \eta_{w\bar{w}} + b(|w|) w \bar{w}}{|w|^2})~.
\ee
Now let us take $w$ to be real, ie $t=0$, $w=\bar{w} = x$. Then $\partial f(|w|) = \bar\partial f(|w|) \vert_{w\in \mathbb{R}}$.
Therefore, we can now take the difference between (\ref{dkL}) and (\ref{dkR}) to get
\be
[\frac{d}{dx}(k_L (x)- k_R(x))]= 0 ~.
\ee
We have thus shown that $k_L- k_R$ is a conserved quantity as we increase $x$, equivalent to the effect of an RG flow towards the infrared.

With an unbalanced $k_L-k_R$, we do not expect a gapped infrared theory. It is however clear that any solution in Einstein scalar theory continues to be an exact solution when the Chern-Simons gauge terms are included in the action. Moreover, in this case, the entanglement entropy formula is sensitive only to the metric and immediately we conclude that the log term is absent!

The resolution of the apparent paradox is that the contribution of the $U(1)$ to the entanglement is invisible to the Ryu-Takayanagi holographic entanglement because the $U(1)$'s contribute to the central charge by an order one effect relative to the large classical contribution of the geometry; this effect which is subleading in the large $N\sim {\ell/G_3}$ limit. To recover the effect of the current algebra, one has to take into account the quantum contribution to the path-integral of the Chern-Simons terms. 

The quantum contribution of the Chern-Simons terms basically boils down to the contribution of the boundary modes.
Let us consider the way it works in pure AdS space.  The quantum contribution of the boundary modes to the action is
given in terms of modular functions.

We are in particular interested in the quantum contribution of the boundary modes to the entanglement entropy. To deal with the replicated geometry, we could play the same trick as in \cite{renyi} and make use of the hyperbolic slicing. The replica index $n$ becomes the temperature of a bulk hyperbolic black hole, and controls the periodicity of the time coordinate. One can compute the boundary partition function using the Cardy formula, where the time direction has periodicity \be \frac{1}{T_n}\equiv\beta_n = 2\pi n~,
\ee and spatial volume is infinite, but has a cutoff that takes the form \be
L = 2\log(\frac{R}{\delta}),
\ee 
where $R$ is the size of the interval, and $\delta$ a short distance cutoff from the boundary of the integral.Together, this gives

\be
\log Z_n= \log (\tr \exp(-\beta H)) = \frac{\pi(c_L+c_R)}{12} \frac{L}{\beta}~,
\ee
and finally the entanglement entropy is given by
\be
S_{EE} =\frac{n}{1-n}\frac{T_n \log Z_n- T_1 \log Z_1}{T_1}= \frac{c_L+c_R}{6}\log(\frac{R}{\delta})~.\ee

For each copy of the gauge field with positive (negative) level, it contributes to a left (right) moving chiral bosonic mode at the boundary. This contribution to the entanglement entropy is thus clearly subleading in $1/N^2$ compared to the gravitational sector.

The point is that since the Chern-Simons terms are topological, it is insensitive to the internal geometry so long as it is regular (in this case it is important to regulate the curvature divergences mentioned in section \ref{sec:general}), and so the only contribution to the quantum action comes from the boundary, which stays the same, unaffected by the holographic RG flow induced by the neutral scalar field $U(1)$. It contributes to a $\log$ term confirming that the infrared is not completely gapped, although these massless modes are not visible in the leading large $N$ limit.

%\section{Marginal gap in Einstein-Scalar theory}\label{app:ES}

\bibliographystyle{utphys}
\bibliography{gapped-bib}

\end{document}